\documentclass[aps,prc,preprint,showpacs,showkeys]{revtex4}
\usepackage{amssymb}
\usepackage{amsmath}
\usepackage{graphicx}
\usepackage{bm}
\usepackage{epsf}

\begin{document}

\title{Influence of the hadronic equation of state on the hadron-quark phase
transition in neutron stars }
\author{F. Yang}
\affiliation{Department of Physics, Nankai University, Tianjin 300071, China}
\author{H. Shen}
\email{songtc@nankai.edu.cn}
\affiliation{Department of Physics, Nankai University, Tianjin 300071, China}

\begin{abstract}
We study the hadron-quark phase transition in the interior of neutron stars.
The relativistic mean field (RMF) theory is adopted to describe the hadronic
matter phase, while the Nambu-Jona-Lasinio (NJL) model is used for the quark matter
phase. The influence of the hadronic equation of state on the phase transition
and neutron star properties are investigated. We find that a neutron star
possesses a large population of hyperons, but it is not dense enough to possess
a pure quark core. Whether the mixed phase of hadronic and quark matter exist
in the core of neutron stars depends on the RMF parameters used.
\end{abstract}



\maketitle

\section{Introduction}

\label{sec:1}

Neutron stars are laboratories for dense matter physics, since they contain
matter in one of the densest forms found in the universe. It is expected
that the deconfinement phase transition may occur in the core of massive
neutron stars~\cite{Weber05}. The study of the hadron-quark phase transition
at high density is of great interest in both nuclear physics and
astrophysics. It has been pointed out by Glendenning~\cite{GL92} that the
hadron-quark phase transition in neutron stars may proceed through a mixed
phase of hadronic and quark matter over a finite range of pressures and
densities according to the Gibbs criteria for phase equilibrium. Such phase
transition has received much attention in neutron star physics~\cite%
{PR97,PR00,PRC99,NPA00,Menezes,Sharma,Lattimer,Burgio,HuangM}.
It is believed that hyperons may appear around twice normal nuclear matter
density through the weak interaction~\cite{PRC96}, which usually occur
earlier than the hadron-quark phase transition. The inclusion of hyperons
in the hadronic phase alters the threshold density and properties of the
mixed phase significantly. In general, the presence of new degrees of freedom,
such as hyperons and quarks, tends to soften the equation of state (EOS) at
high density and lower the maximum mass of neutron stars.

To study the hadron-quark phase transition, we need models to describe
hadronic matter and quark matter. Unfortunately, there is no single model
which can be used to describe both phases and the dynamic process of the
phase transition. We have to use different approaches for the description of
the two phases, then perform the Glendenning construction for the
charge-neutral mixed phase where both hadronic and quark phases coexist~\cite%
{GL92}. In this work, we adopt the relativistic mean field (RMF) theory to
describe the hadronic matter phase, while the Nambu-Jona-Lasinio (NJL) model
is used for the quark matter phase. The choice of the NJL model is motivated
by the fact that this model can successfully reproduce many aspects of
quantum chromodynamics such as the nonperturbative vacuum structure and
dynamical breaking of chiral symmetry~\cite{NJL1,NJL2,NJL3}. We adopt a
three-flavor version of the NJL model to describe the quark matter phase~%
\cite{NJL1}. With a definite EOS for quark matter based on the NJL model, we
examine the influence of the hadronic EOS on the hadron-quark phase
transition and neutron star properties.

We use the RMF theory to describe the hadronic matter phase. The RMF theory
has been successfully and widely used for the description of nuclear
matter and finite nuclei~\cite{Serot86,Ring90,Toki96,Ren02,Shen06}. It has
also been applied to provide the EOS of dense matter for the use in
supernovae and neutron stars~\cite{Shen98,Shen02}. In the RMF approach,
baryons interact through the exchange of scalar and vector mesons. The
meson-nucleon coupling constants are generally determined by fitting to some
nuclear matter properties or ground-state properties of finite nuclei.
However, there are large uncertainties in the meson-hyperon couplings due to
limited experimental data. One can use the coupling constants derived from
the quark model, or those values constrained by reasonable hyperon
potentials. The two additional strange mesons, $\sigma
^{\ast }$ and $\phi $, were originally introduced in order to obtain the
strong attractive hyperon-hyperon ($YY$) interaction deduced from the
earlier measurement~\cite{PRL93}. A recent observation of the double-$%
\Lambda $ hypernucleus $_{\Lambda \Lambda }^{6}\mathrm{He}$, called the
Nagara event~\cite{Nagara}, has had a significant impact on strangeness
nuclear physics. The Nagara event provides unambiguous identification of $%
_{\Lambda \Lambda }^{6}\mathrm{He}$ production with precise $\Lambda \Lambda
$ binding energy value $B_{\Lambda \Lambda }=7.25\pm 0.19_{-0.11}^{+0.18}\;%
\mathrm{MeV}$, which suggests that the effective $\Lambda \Lambda $
interaction should be considerably weaker ($\triangle B_{\Lambda \Lambda
}\sim 1\;\mathrm{MeV}$) than that deduced from the earlier measurement ($%
\triangle B_{\Lambda \Lambda }\sim 5\;\mathrm{MeV}$). The weak $YY$
interaction suggested by the Nagara event has been used to reinvestigate the
properties of multistrange systems, and it has been found that the change
of $YY$ interactions affects the properties of strange hadronic matter
dramatically~\cite{PRC03s,JPG04s,JPG05}. In order to examine the influence
of the hadronic EOS on the hadron-quark phase transition, we employ two
successful parameter sets of the RMF model, NL3~\cite{NL3} and TM1~\cite{TM1}%
, which have been widely used for the description of nuclear matter and
finite nuclei including unstable nuclei. For each parameter set of the
nucleonic sector, we consider two cases of hyperon-hyperon interactions, the
weak and strong $YY$ interactions. By comparing the results with
different parametrizations in the RMF model, we evaluate how sensitive the
hadron-quark phase transition is to the hadronic EOS used in the calculation.

For a comprehensive description of neutron stars, we need not only the EOS
at high density for the interior region but also the EOS for the inner and
outer crusts, where the density is low and heavy nuclei exist. For the
nonuniform matter at low density, we adopt a relativistic EOS based on the
RMF theory with a local density approximation~\cite{Shen98,Shen02}. The
nonuniform matter is modelled to be composed of a lattice of spherical
nuclei immersed in an electron gas with or without free neutrons dripping
out of nuclei. As the density increases, heavy nuclei dissolve and the
optimal state is a uniform matter consisting of neutrons, protons, and
leptons (electrons and muons) in $\beta $ equilibrium. The low density EOS
is therefore matched to an EOS of uniform nuclear matter at around $10^{14}$
g/cm$^{3}$~\cite{Shen02}. At higher densities, some additional degrees of
freedom such as hyperons and quarks may occur, and the hadron-quark phase
transition could proceed through a mixed phase of hadronic and quark
matter~\cite{PRC99,NPA00,Menezes,Sharma,Lattimer,Burgio,HuangM}.
Applying the EOS of neutron star matter over a wide density range,
we study the neutron star properties by solving the
Tolman-Oppenheimer-Volkoff equation, and examine whether or
not the deconfined quark phase can exist in the core of neutron stars.

This paper is arranged as follows. In Sec.~\ref{sec:2}, we discuss the EOS
for hadronic matter in the RMF theory, and the parameters used in the calculation.
In Sec.~\ref{sec:3}, the NJL model is used for the description of quark
matter. In Sec.~\ref{sec:4}, we investigate the hadron-quark phase
transition of neutron star matter, and examine the influence of the hadronic
EOS. We present in Sec.~\ref{sec:5} the properties of neutron stars.
Section~\ref{sec:6} is devoted to a summary.


\section{Hadronic Phase}

\label{sec:2}

To describe the hadronic matter phase, we adopt the relativistic mean field
(RMF) theory, in which baryons interact via the exchange of mesons. The
baryons considered in this work are nucleons ($p$ and $n$) and hyperons ($%
\Lambda $, $\Sigma $, and $\Xi $). The exchanged mesons include isoscalar
scalar and vector mesons ($\sigma $ and $\omega $), isovector vector meson ($%
\rho $), and two additional hidden-strangeness mesons ($\sigma ^{\ast }$ and
$\phi $). For neutron star matter consisting of a neutral mixture of baryons
and leptons in $\beta $ equilibrium, we start from the effective Lagrangian%
\begin{eqnarray}
\mathcal{L}_{RMF} &=&\sum_{B}\bar{\psi}_{B}\left[ i\gamma _{\mu }\partial
^{\mu }-m_{B}-g_{\sigma B}\sigma -g_{\sigma ^{\ast }B}\sigma ^{\ast
}-g_{\omega B}\gamma _{\mu }\omega ^{\mu }\right.  \nonumber \\
&&\left. -g_{\phi B}\gamma _{\mu }\phi ^{\mu }-g_{\rho B}\gamma _{\mu }\tau
_{i}\rho _{i}^{\mu }\right] \psi _{B}+\frac{1}{2}\partial _{\mu }\sigma
\partial ^{\mu }\sigma -\frac{1}{2}m_{\sigma }^{2}\sigma ^{2}  \nonumber \\
&&-\frac{1}{3}g_{2}\sigma ^{3}-\frac{1}{4}g_{3}\sigma ^{4}-\frac{1}{4}W_{\mu
\nu }W^{\mu \nu }+\frac{1}{2}m_{\omega }^{2}\omega _{\mu }\omega ^{\mu }
\nonumber \\
&&+\frac{1}{4}c_{3}\left( \omega _{\mu }\omega ^{\mu }\right) ^{2}-\frac{1}{4%
}R_{i\mu \nu }R_{i}^{\mu \nu }+\frac{1}{2}m_{\rho }^{2}\rho _{i\mu }\rho
_{i}^{\mu }  \nonumber \\
&&+\frac{1}{2}\partial _{\mu }\sigma ^{\ast }\partial ^{\mu }\sigma ^{\ast }-%
\frac{1}{2}m_{\sigma ^{\ast }}^{2}\sigma ^{\ast 2}-\frac{1}{4}S_{\mu \nu
}S^{\mu \nu }+\frac{1}{2}m_{\phi }^{2}\phi _{\mu }\phi ^{\mu }  \nonumber \\
&&+\sum_{l}\bar{\psi}_{l}\left[ i\gamma _{\mu }\partial ^{\mu }-m_{l}\right]
\psi _{l},  \label{eq:Lrmf}
\end{eqnarray}%
where the sum on $B$ runs over the baryon octet ($p$, $n$, $\Lambda $, $%
\Sigma ^{+}$, $\Sigma ^{0}$, $\Sigma ^{-}$, $\Xi ^{0}$, $\Xi ^{-}$),
and the sum on $l$ is over electrons and muons ($e^{-}$ and $\mu
^{-}$). In the RMF model, the meson fields are treated as classical
fields, and the field operators are replaced by their expectation
values. The meson field equations in uniform matter have the
following form:
\begin{eqnarray}
&&m_{\sigma }^{2}\sigma +g_{2}\sigma ^{2}+g_{3}\sigma ^{3}=-\sum_{B}\frac{%
g_{\sigma B}}{\pi ^{2}}\int_{0}^{k_{F}^{B}}\frac{m_{B}^{\ast }}{\sqrt{%
k^{2}+m_{B}^{\ast 2}}}k^{2}dk,  \label{eq:s} \\
&&m_{\omega }^{2}\omega +c_{3}\omega ^{3}=\sum_{B}\frac{g_{\omega B}\left(
k_{F}^{B}\right) ^{3}}{3\pi ^{2}},  \label{eq:w} \\
&&m_{\rho }^{2}\rho =\sum_{B}\frac{g_{\rho B}\tau _{3B}\left(
k_{F}^{B}\right) ^{3}}{3\pi ^{2}},  \label{eq:r} \\
&&m_{\sigma ^{\ast }}^{2}\sigma ^{\ast }=-\sum_{B}\frac{g_{\sigma ^{\ast }B}%
}{\pi ^{2}}\int_{0}^{k_{F}^{B}}\frac{m_{B}^{\ast }}{\sqrt{k^{2}+m_{B}^{\ast
2}}}k^{2}dk,  \label{eq:ss} \\
&&m_{\phi }^{2}\phi =\sum_{B}\frac{g_{\phi B}\left( k_{F}^{B}\right) ^{3}}{%
3\pi ^{2}},  \label{eq:ws}
\end{eqnarray}%
where $\sigma =\left\langle \sigma \right\rangle ,$ $\omega =\left\langle
\omega ^{0}\right\rangle ,$ $\rho =\left\langle \rho ^{30}\right\rangle ,$ $%
\sigma ^{\ast }=\left\langle \sigma ^{\ast }\right\rangle ,$ and $\phi
=\left\langle \phi ^{0}\right\rangle $ are the nonvanishing expectation
values of meson fields in neutron star matter. $m_{B}^{\ast
}=m_{B}+g_{\sigma B}\sigma +g_{\sigma ^{\ast }B}\sigma ^{\ast }$ is the
effective mass of the baryon species $B$, and $k_{F}^{B}$ is the Fermi
momentum.

In this work, we employ two successful parameter sets of the RMF model, NL3
and TM1, as listed in Table~\ref{tab:1}. These parameters have been
determined by fitting to some ground-state properties of finite nuclei, and
they can provide good description of nuclear matter and finite nuclei
including unstable nuclei~\cite{NL3,TM1}. As for the meson-hyperon
couplings, we take the naive quark model values for the vector coupling
constants,
\begin{eqnarray}
&&\frac{1}{3}g_{\omega N}=\frac{1}{2}g_{\omega \Lambda }=\frac{1}{2}%
g_{\omega \Sigma }=g_{\omega \Xi },  \nonumber \\
&&g_{\rho N}=\frac{1}{2}g_{\rho \Sigma }=g_{\rho \Xi },\ \ g_{\rho \Lambda
}=0, \nonumber \\
&&2g_{\phi \Lambda }=2g_{\phi \Sigma }=g_{\phi \Xi }=-\frac{2\sqrt{2}}{3}%
g_{\omega N},\ \ g_{\phi N}=0.
\end{eqnarray}%
The scalar coupling constants are chosen to give reasonable hyperon
potentials. The potential depth of the hyperon species $i$ in matter of the
baryon species $j$ is denoted by $U_{i}^{\left( j\right) }$, and we use
$U_{\Lambda }^{\left(N\right) }=-28$ MeV,
$U_{\Sigma  }^{\left(N\right) }=+30$ MeV, and
$U_{\Xi     }^{\left(N\right) }=-18$ MeV~\cite{PRC88,PRC00,PRC04}
to determine the scalar coupling constants
$g_{\sigma \Lambda }$, $g_{\sigma \Sigma }$, and $g_{\sigma \Xi }$,
respectively. The hyperon couplings to strange meson $\sigma^{\ast }$ are
restricted by the relation $U_{\Xi }^{\left(\Xi\right) }\simeq U_{\Lambda
}^{\left(\Xi\right) }\simeq 2U_{\Xi}^{\left(\Lambda\right) }\simeq
2U_{\Lambda }^{\left(\Lambda\right) }$ obtained in Ref.~\cite{ANN94}.
We consider two cases of hyperon-hyperon ($YY$) interactions.
The strong $YY$ interaction deduced from the earlier measurement~\cite{PRL93}
suggests $U_{\Lambda }^{(\Lambda )}\simeq -20\ \mathrm{MeV,}$ while the weak
$YY$ interaction implied by the Nagara event suggests
$U_{\Lambda}^{(\Lambda )}\simeq -5\ \mathrm{MeV}$~\cite{PRC03s,JPG04s,JPG05}.
In Table~ \ref{tab:2}, we present the meson-hyperon couplings
determined by these hyperon potentials. The hyperon masses are taken to be
$m_{\Lambda }=1115.7\ \text{MeV}$, $m_{\Sigma }=1193.1\ \text{MeV}$,
and $m_{\Xi }=1318.1\ \text{MeV}$, while the strange meson masses are
$m_{\sigma^{\ast }}=975\ \text{MeV}$ and $m_{\phi }=1020\ \text{MeV}$.

For neutron star matter consisting of a neutral mixture of baryons and
leptons, the $\beta $ equilibrium conditions without trapped neutrinos are
given by%
\begin{eqnarray}
&&\mu _{p}=\mu _{\Sigma ^{+}}=\mu _{n}-\mu _{e},  \label{eq:beta11} \\
&&\mu _{\Lambda }=\mu _{\Sigma ^{0}}=\mu _{\Xi ^{0}}=\mu _{n},
\label{eq:beta12} \\
&&\mu _{\Sigma ^{-}}=\mu _{\Xi ^{-}}=\mu _{n}+\mu _{e},  \label{eq:beta13} \\
&&\mu _{\mu }=\mu _{e},  \label{eq:beta14}
\end{eqnarray}%
where $\mu _{i}$ is the chemical potential of species $i$. At zero
temperature the chemical potentials of baryons and leptons are expressed by%
\begin{eqnarray}
&&\mu _{B}=\sqrt{{k_{F}^{B}}^{2}+m_{B}^{\ast 2}}+g_{\omega B}\omega +g_{\phi
B}\phi +g_{\rho B}\tau _{3B}\rho ,  \label{eq:mub} \\
&&\mu _{l}=\sqrt{{k_{F}^{l}}^{2}+m_{l}^{2}}.  \label{eq:mul}
\end{eqnarray}%
The charge neutrality condition is given by
\begin{eqnarray}
n_{p}+n_{\Sigma ^{+}}=n_{e}+n_{\mu }+n_{\Sigma ^{-}}+n_{\Xi ^{-}},
\end{eqnarray}
where $n_{i}=\left( k_{F}^{i}\right) ^{3}/(3\pi ^{2})$ is the number density
of species $i$. We can solve the coupled equations self-consistently at a
given baryon density $n_{B}=n_{p}+n_{n}+n_{\Lambda }+n_{\Sigma
^{+}}+n_{\Sigma ^{0}}+n_{\Sigma ^{-}}+n_{\Xi ^{0}}+n_{\Xi ^{-}}$. The total
energy density and pressure of neutron star matter are written by
\begin{eqnarray}
\varepsilon _{HP} &=&\sum_{B}\frac{1}{\pi ^{2}}\int_{0}^{k_{F}^{B}}\sqrt{%
k^{2}+m_{B}^{\ast 2}}\ k^{2}dk+\frac{1}{2}m_{\sigma }^{2}\sigma ^{2}+\frac{1%
}{3}g_{2}\sigma ^{3}+\frac{1}{4}g_{3}\sigma ^{4}  \nonumber \\
&&+\frac{1}{2}m_{\omega }^{2}\omega ^{2}+\frac{3}{4}c_{3}\omega ^{4}+\frac{1%
}{2}m_{\rho }^{2}\rho ^{2}+\frac{1}{2}m_{\sigma ^{\ast }}^{2}\sigma ^{\ast
2}+\frac{1}{2}m_{\phi }^{2}\phi ^{2}  \nonumber \\
&&+\sum_{l}\frac{1}{\pi ^{2}}\int_{0}^{k_{F}^{l}}\sqrt{k^{2}+m_{l}^{2}}\
k^{2}dk,  \label{eq:e1}
\end{eqnarray}%
\begin{eqnarray}
P_{HP} &=&\frac{1}{3}\sum_{B}\frac{1}{\pi ^{2}}\int_{0}^{k_{F}^{B}}\frac{%
k^{4}\ dk}{\sqrt{k^{2}+m_{B}^{\ast 2}}}-\frac{1}{2}m_{\sigma }^{2}\sigma
^{2}-\frac{1}{3}g_{2}\sigma ^{3}-\frac{1}{4}g_{3}\sigma ^{4}  \nonumber \\
&&+\frac{1}{2}m_{\omega }^{2}\omega ^{2}+\frac{1}{4}c_{3}\omega ^{4}+\frac{1%
}{2}m_{\rho }^{2}\rho ^{2}-\frac{1}{2}m_{\sigma ^{\ast }}^{2}\sigma ^{\ast
2}+\frac{1}{2}m_{\phi }^{2}\phi ^{2}  \nonumber \\
&&+\frac{1}{3}\sum_{l}\frac{1}{\pi ^{2}}\int_{0}^{k_{F}^{l}}\frac{k^{4}\ dk}{%
\sqrt{k^{2}+m_{l}^{2}}}.  \label{eq:p1}
\end{eqnarray}


\section{Quark Phase}

\label{sec:3}

In this section, we adopt a three-flavor version of the NJL model to
describe the deconfined quark phase. The Lagrangian is given by%
\begin{eqnarray}
\mathcal{L}_{NJL} &=&\bar{q}\left( i\gamma _{\mu }\partial ^{\mu
}-m^{0}\right) q+G\sum_{a=0}^{8}\left[ \left( \bar{q}\lambda _{a}q\right)
^{2}+\left( \bar{q}i\gamma _{5}\lambda _{a}q\right) ^{2}\right]  \nonumber \\
&&-K\left\{ \det \left[ \bar{q}\left( 1+\gamma _{5}\right) q\right] +\det %
\left[ \bar{q}\left( 1-\gamma _{5}\right) q\right] \right\} ,
\label{eq:Lnjl}
\end{eqnarray}%
where $q$ denotes a quark field with three flavors $\left( N_{f}=3\right) $
and three colors $\left( N_{c}=3\right) $. $m^{0}=$diag$\left(
m_{u}^{0},m_{d}^{0},m_{s}^{0}\right) $ is the current quark mass matrix, and
we assume isospin symmetry $m_{u}^{0}=m_{d}^{0}\equiv m_{q}^{0}$. The
coupling constants $G$ and $K$ have dimension energy$^{-2}$ and energy$^{-5}$%
, respectively. The model has five parameters, namely, the current quark
masses $m_{q}^{0}$ and $m_{s}^{0}$, the coupling constants $K$ and $G$, and
the momentum cutoff $\Lambda $. In the present calculation, we employ the
parameters given in Ref.~\cite{NJL96}, $m_{q}^{0}=5.5\ \text{MeV}$, $%
m_{s}^{0}=140.7\ \text{MeV}$, $\Lambda =602.3\ \text{MeV}$, $G\Lambda
^{2}=1.835$, and $K\Lambda ^{5}=12.36$. These parameters have been
determined by fitting $f_{\pi }$, $m_{\pi }$, $m_{K}$, and $m_{\eta
^{^{\prime }}}$ to their empirical values, while the mass of the $\eta $%
-meson is underestimated by about 6\%~\cite{NJL1}.

In the NJL model, the quark gets constituent quark mass by spontaneous
chiral symmetry breaking. The constituent quark mass in vacuum $m_{i}$ is
much larger than the current quark mass $m_{i}^{0}$. In the quark matter
at high density, the constituent quark mass $m_{i}^{\ast }$ becomes
approximately the same as $m_{i}^{0}$, which reflects the restoration of
chiral symmetry. Within the mean-field approximation, $m_{i}^{\ast }$ is
obtained by solving the gap equation%
\begin{equation}
m_{i}^{\ast }=m_{i}^{0}-4G\langle \bar{q}_{i}q_{i}\rangle +2K\langle \bar{q}%
_{j}q_{j}\rangle \langle \bar{q}_{k}q_{k}\rangle ,  \label{eq:gap}
\end{equation}%
with ($i,j,k$) being any permutation of ($u,d,s$). The quark condensate $%
C_{i}=\left\langle \bar{q}_{i}q_{i}\right\rangle $ is given by
\begin{equation}
C_{i}=-\frac{3}{{\pi }^{2}}{\int_{k_{F}^{i}}^{\Lambda }\frac{m_{i}^{\ast }}{%
\sqrt{k^{2}+m_{i}^{\ast 2}}}}k^{2}dk,
\end{equation}
where $k_{F}^{i}$ denotes the Fermi momentum of the quark flavor $i$,
which is connected with the number density $n_{i}$ and the
chemical potential $\mu _{i}$ via
\begin{eqnarray}
n_{i} &=&\frac{\left( k_{F}^{i}\right) ^{3}}{\pi ^{2}}, \\
\mu _{i} &=&\sqrt{{k_{F}^{i}}^{2}+{m_{i}^{\ast }}^{2}}.
\end{eqnarray}%
The energy density of the quark system is given by
\begin{eqnarray}
\varepsilon _{NJL} &=&\sum_{i=u,d,s}\left[ -\frac{3}{\pi ^{2}}%
\int_{k_{F}^{i}}^{\Lambda }\sqrt{k^{2}+m_{i}^{\ast 2}}\ k^{2}dk\right]
\nonumber \\
&&+2G\left( C_{u}^{2}+C_{d}^{2}+C_{s}^{2}\right) -4K{C_{u}}{C_{d}}{C_{s}}%
-\varepsilon _{0},  \label{eq:eNJL}
\end{eqnarray}%
where $\varepsilon _{0}$ is introduced to ensure $\varepsilon _{NJL}=0$ in
the vacuum.

For the quark matter consisting of a neutral mixture of quarks ($u$, $d$,\
and $s$) and leptons ($e$ and $\mu $) in $\beta $ equilibrium, the charge
neutrality condition is expressed as%
\begin{equation}
\frac{2}{3}n_{u}-\frac{1}{3}\left( n_{d}+n_{s}\right) -n_{e}-n_{\mu }=0,
\label{eq:charge2}
\end{equation}%
the $\beta $ equilibrium conditions are given by%
\begin{eqnarray}
\mu _{s} &=&\mu _{d}=\mu _{u}+\mu _{e},  \label{eq:beta21} \\
\mu _{\mu } &=&\mu _{e}.  \label{eq:beta22}
\end{eqnarray}%
The coupled equations can be solved self-consistently at a given baryon
density $n_{B}=\left( n_{u}+n_{d}+n_{s}\right) /3$. The total energy density
and pressure including the contributions from both quarks and leptons are
given by
\begin{eqnarray}
\varepsilon _{QP} &=&\varepsilon _{NJL}+\sum_{l=e,\mu }\frac{1}{\pi ^{2}}%
\int_{0}^{k_{F}^{l}}\sqrt{k^{2}+m_{l}^{2}}\ k^{2}dk,  \label{eq:e2} \\
P_{QP} &=&\sum_{i=u,d,s,e,\mu }n_{i}\mu _{i}-\varepsilon _{QP}.
\label{eq:p2}
\end{eqnarray}


\section{Hadron-Quark Phase Transition}

\label{sec:4}

In this section, we study the hadron-quark phase transition which may occur
in the core of massive neutron stars. It has been discussed extensively in
the literature that a mixed phase of hadronic and quark matter could exist
over a finite range of pressures and densities according to the Gibbs
criteria for phase equilibrium. In the mixed phase, the local charge
neutrality condition is replaced by a global one. This means that both
hadronic and quark matter are allowed to be separately charged. The
condition of global charge neutrality is expressed as
\begin{equation}
\chi n_{c}^{QP}+\left( 1-\chi \right) n_{c}^{HP}=0,  \label{eq:charge3}
\end{equation}%
where $\chi $ is the volume fraction occupied by quark matter in the mixed
phase, which increases from $\chi =0$ in the pure hadronic
phase to $\chi =1$ in the pure quark phase. $n_{c}^{HP}$ and $n_{c}^{QP}$
denote the charge densities of hadronic phase and quark phase, respectively.
Without the constraint of local charge neutrality, we impose that the two
phases are in weak equilibrium and described by two independent chemical
potentials ($\mu_{n}$, $\mu_{e}$). The Gibbs condition for phase
equilibrium at zero temperature is then given by
\begin{equation}
P_{HP}\left( \mu _{n},\mu _{e}\right) =P_{QP}\left( \mu _{n},\mu _{e}\right).
\label{eq:p3}
\end{equation}%
Using Eq.~(\ref{eq:p3}) we can calculate the equilibrium chemical potentials
of the mixed phase where $P_{HP}=P_{QP}=P_{MP}$ holds. The energy density
and the baryon density in the mixed phase are given by
\begin{equation}
\varepsilon _{MP}=\chi \varepsilon _{QP}+\left( 1-\chi \right) \varepsilon
_{HP},  \label{eq:e3}
\end{equation}
and
\begin{equation}
n_{B}^{MP}=\chi n_{B}^{QP}+\left( 1-\chi \right) n_{B}^{HP}.  \label{eq:r3}
\end{equation}

\begin{figure}[tbh]
\begin{center}
\includegraphics[bb=10 10 540 560, width=8.6 cm,clip]{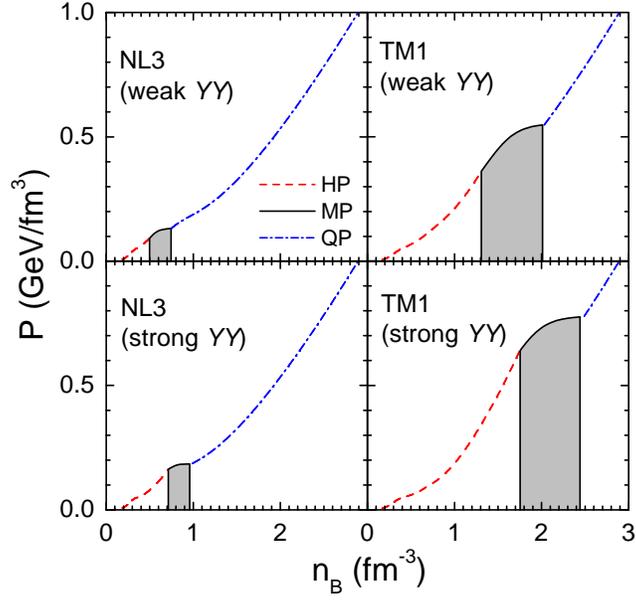}
\end{center}
\caption{(Color online) The pressure $P$ as a function of the baryon density $n_B$.
The shaded regions correspond to the mixed phase (MP). The dashed and
dot-dashed lines show the pressures of hadronic phase (HP) and quark
phase (QP), respectively.}
\label{fig:1}
\end{figure}

\begin{figure}[tbh]
\begin{center}
\includegraphics[bb=10 10 520 550, width=8.6 cm,clip]{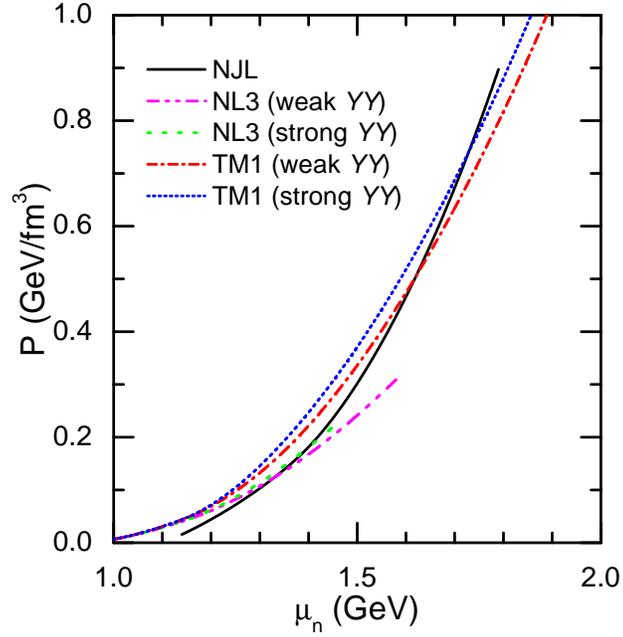}
\end{center}
\caption{(Color online) The pressure $P$ as a function of the neutron chemical
potential $\mu _{n}$ for hadronic and quark matter with local charge
neutrality. The crossing of the hadronic EOS with the quark EOS (NJL)
marks the transition point between the charge neutral hadronic matter
and quark matter.}
\label{fig:2}
\end{figure}

We show in Fig.~\ref{fig:1} the possible phase structure of neutron star
matter using the RMF model for the hadronic phase and the NJL model for the quark
phase. To examine the influence of the hadronic EOS on the hadron-quark
phase transition, we employ two successful parameter sets of the RMF model,
NL3~\cite{NL3} and TM1~\cite{TM1}, with both the weak and strong $YY$
interactions. The shaded regions correspond to the mixed phase. It is shown
that a pure hadronic phase is favored at low density. The mixed phase
appears at the critical density $n_{B}^{(1)}$ where the pressure of the pure
hadronic phase becomes to be lower than the pressure of the mixed phase. The
fraction of quark matter $\chi $ increases with increasing density in the
mixed phase. It turns to be a pure quark phase at the critical density $%
n_{B}^{(2)}$ where the pressure of the pure quark phase is above the
pressure of the mixed phase. The critical densities $n_{B}^{(1)}$ and $%
n_{B}^{(2)}$ are sensitive to the RMF parameters used, and we get $%
n_{B}^{(1)}\simeq 0.50\ \mathrm{{fm}^{-3}}$ ($0.71\ \mathrm{{fm}^{-3}}$) and
$n_{B}^{(2)}\simeq 0.75\ \mathrm{{fm}^{-3}}$ ($0.98\ \mathrm{{fm}^{-3}}$)
for the NL3 parameter set with the weak (strong) $YY$ interaction,
while $n_{B}^{(1)}\simeq 1.31\ \mathrm{{fm}^{-3}}$ ($1.75\ \mathrm{{fm}^{-3}}
$) and $n_{B}^{(2)}\simeq 2.03\ \mathrm{{fm}^{-3}}$ ($2.49\ \mathrm{{fm}^{-3}%
}$) are obtained for the TM1 cases. It is found that the hadron-quark phase
transition occurs at lower densities in the NL3 model than in the TM1 model,
and in general the weak $YY$ interaction leads to an earlier appearance
of the mixed phase. In order to estimate the influence of hadronic EOS on
the deconfinement phase transition, we plot in Fig.~\ref{fig:2} the hadronic
EOS in the different cases and the quark EOS in the NJL model with local charge
neutrality as a function of the neutron chemical potential $\mu _{n}$.
The crossing of the hadronic EOS with the quark EOS marks the transition
point between the charge neutral hadronic matter and quark matter.
It is seen that the NL3 model favors the phase transition at a lower $\mu_{n}$.
A more realistic treatment of the phase transition is to release the constraint
of local charge neutrality, which leads to the existence of the mixed phase of
charged hadronic and quark matter over a finite range of pressures and
densities as shown in Fig.~\ref{fig:1}. In general, a harder hadronic EOS
also favors an earlier appearance of the mixed phase.

\begin{figure}[tbh]
\begin{center}
\includegraphics[bb=10 60 540 560, width=8.6 cm,clip]{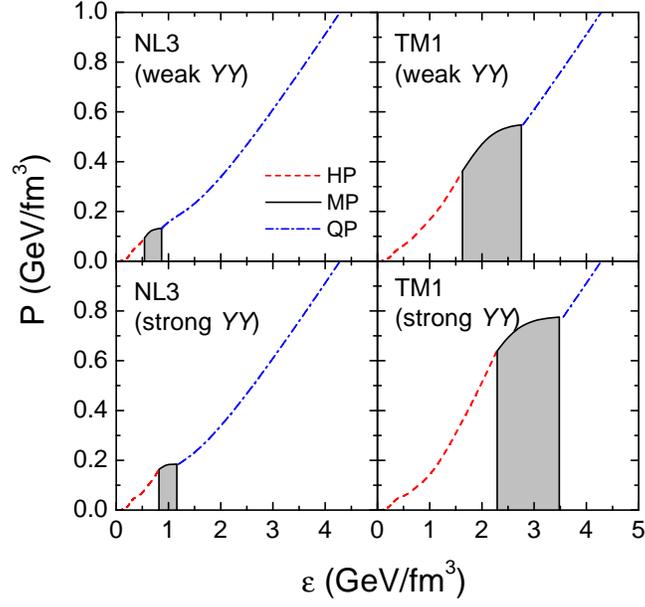}
\end{center}
\caption{(Color online) The full EOS of neutron star matter in the form of pressure $P$
versus energy density $\varepsilon$. The shaded regions correspond to
the mixed phase (MP). The dashed and dot-dashed lines show the pressures
of hadronic phase (HP) and quark phase (QP), respectively.}
\label{fig:3}
\end{figure}

In Fig.~\ref{fig:3} we plot the full EOS in the form $P=P(\varepsilon )$,
which consists of three parts: (a) the charge neutral hadronic matter phase
at low density (b) the mixed phase of charged hadronic and quark matter (c)
the charge neutral quark matter phase at high density. The mixed phase part
of the EOS is shaded gray, where the pressure varies continuously. It is
shown that the onset and width of the mixed phase depend on the RMF
parameters used in the calculation. The NL3 model leads to earlier
appearance of the mixed phase than the TM1 model, and the weak $YY$
interaction favors earlier onset of the mixed phase than the strong $YY$
interaction. This is mainly because that a harder hadronic EOS prefers an
earlier hadron-quark phase transition. By comparing the results of
different cases, we can see the influence of the hadronic EOS on the
hadron-quark phase transition.

\begin{figure}[tbh]
\begin{center}
\includegraphics[bb=10 10 650 660, width=8.6 cm,clip]{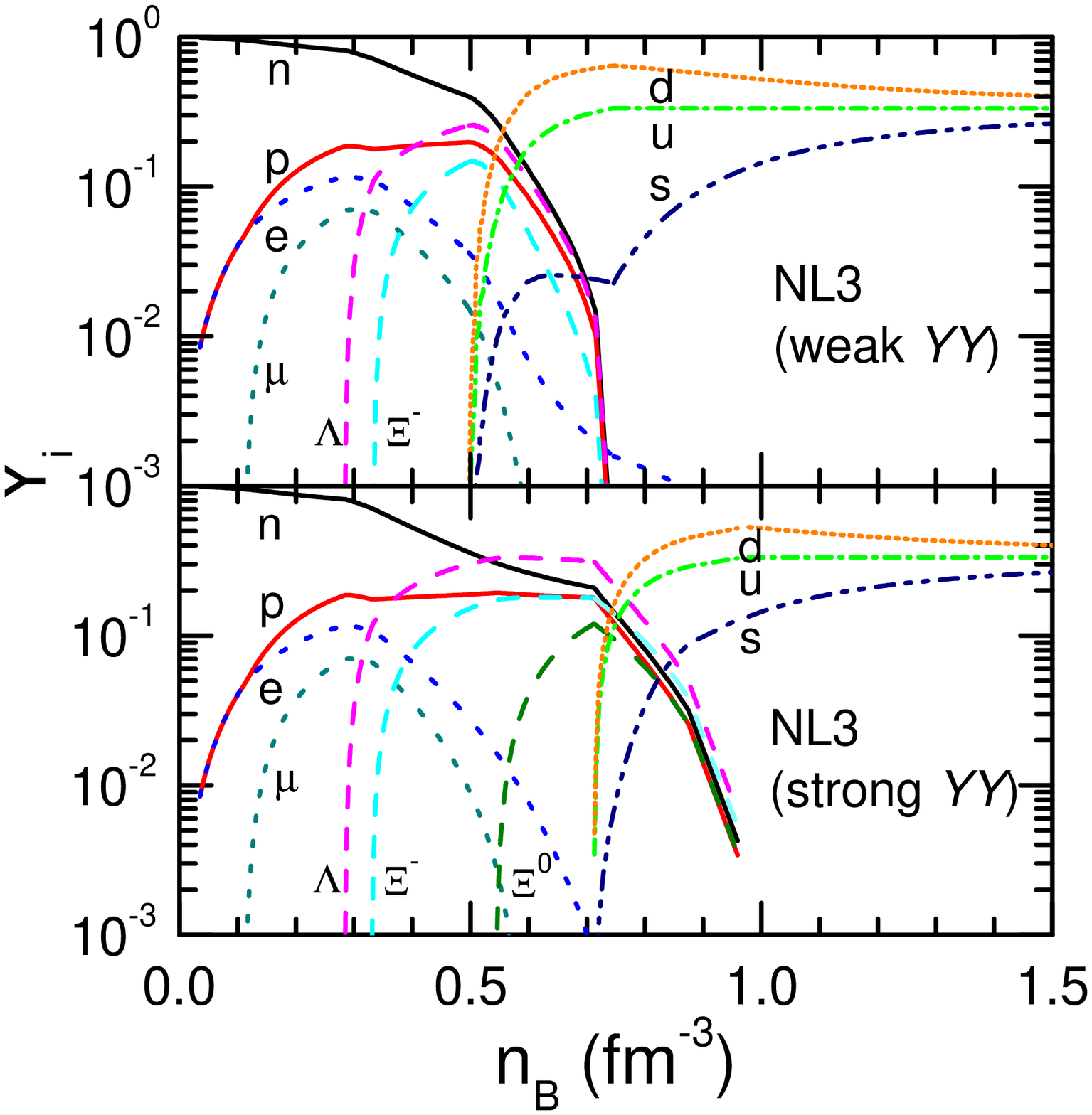}
\end{center}
\caption{(Color online) The particle fraction $Y_{i}=n_{i}/n_{B}$ as a function of
the total baryon density $n_{B}$ for the NL3 model.}
\label{fig:4}
\end{figure}

\begin{figure}[tbh]
\begin{center}
\includegraphics[bb=10 270 580 790, width=8.6 cm,clip]{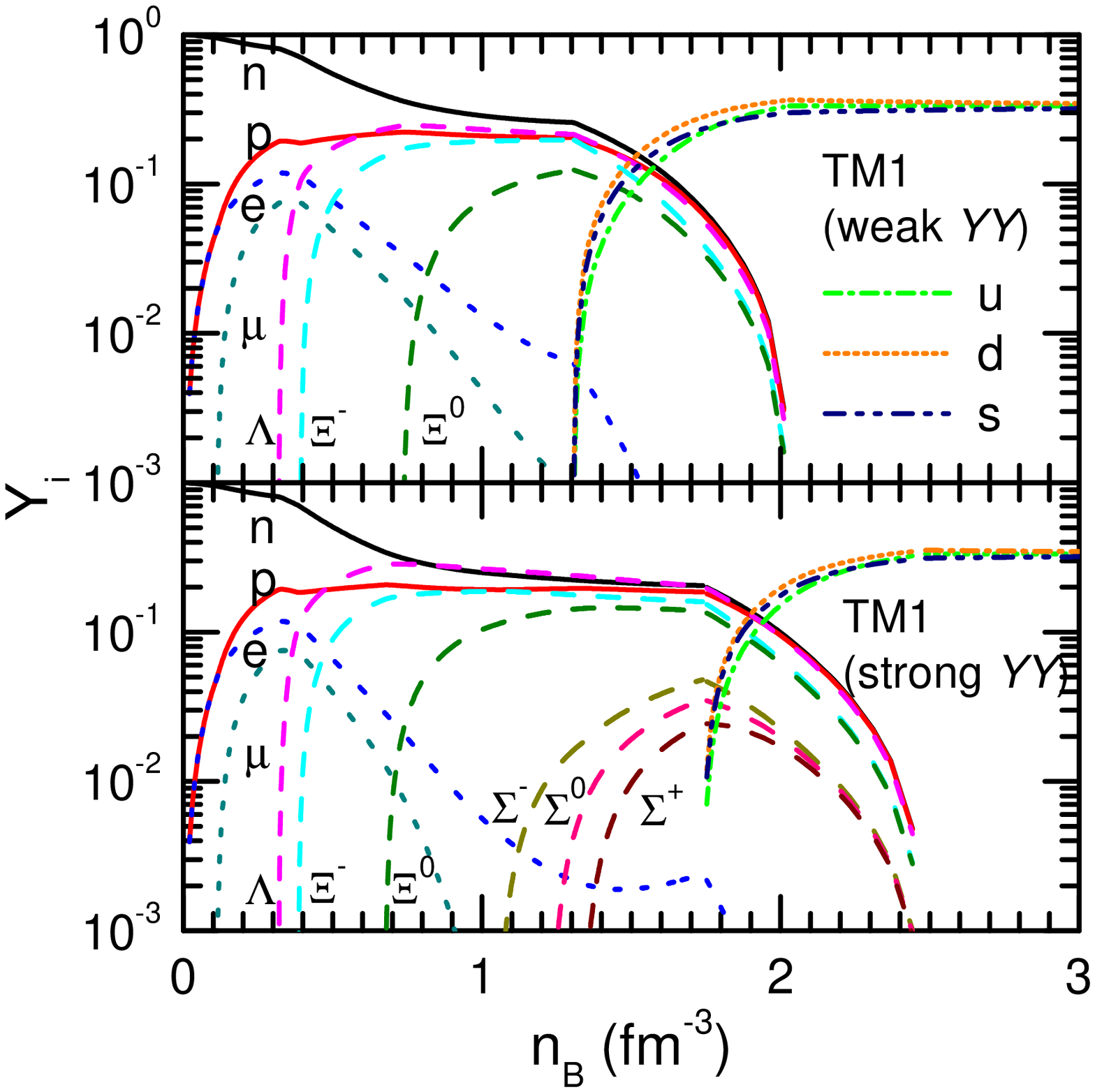}
\end{center}
\caption{(Color online) The particle fraction $Y_{i}=n_{i}/n_{B}$ as a function of
the total baryon density $n_{B}$ for the TM1 model.}
\label{fig:5}
\end{figure}

We present in Figs.~\ref{fig:4} and \ref{fig:5} the particle fraction $%
Y_{i}=n_{i}/n_{B}$ as a function of the total baryon density $n_{B}$. At low
densities the fractions $Y_{p}$, $Y_{e}$, and$\ Y_{\mu }$ increase with
increasing density. When the Fermi energy of nucleons exceeds the rest mass
of hyperons, the conversion of nucleons to hyperons is energetically
favorable, and it can relieve the Fermi pressure of nucleons. The fraction
of hyperons increases with increasing density before the mixed phase occurs.
Quarks appear at the critical density $n_{B}^{(1)}$, then the fractions $%
Y_{u}$, $Y_{d}$, and$\ Y_{s}$ increase rapidly with increasing density. The
hadronic matter completely disappears at the critical density $n_{B}^{(2)}$
where the pure quark phase occurs. At extremely high density,
$Y_{u}\sim Y_{d}\sim Y_{s}\sim 1/3$ due to the restoration of
chiral symmetry. It is shown that the composition of neutron star matter
depends on the RMF parameters used in the calculation.


\section{Neutron stars properties}

\label{sec:5}

In this section, we investigate the properties of neutron stars by solving
the Tolman-Oppenheimer-Volkoff equation with the EOS over a wide density
range. For the nonuniform matter at low density, which exists in the inner
and outer crusts of neutron stars, we adopt a relativistic EOS based on the
RMF theory with a local density approximation~\cite{Shen98,Shen02}. The
nonuniform matter is modelled to be composed of a lattice of spherical
nuclei immersed in an electron gas with or without free neutrons dripping
out of nuclei. The low density EOS is matched to the EOS of uniform hadronic
matter at the density where they have equal pressures. The pure hadronic
phase ends at the critical density $n_{B}^{(1)}$, and the pure quark phase
starts at the critical density $n_{B}^{(2)}$. The values of these critical
densities depend on the RMF parameters used. The neutron star properties
are mainly determined by the EOS at high density. We calculate neutron star
profiles in order to examine whether or not the deconfined quark phase can
exist in the core of neutron stars.

\begin{figure}[tbh]
\begin{center}
\includegraphics[bb=20 20 530 550, width=8.6 cm,clip]{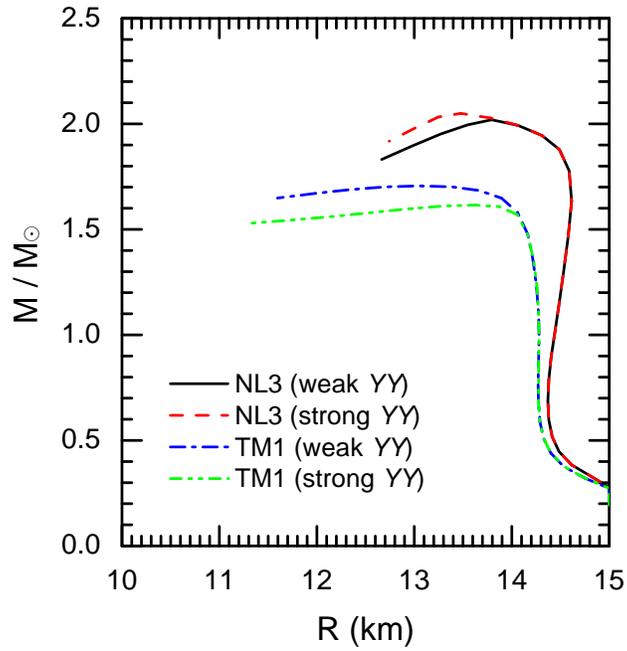}
\end{center}
\caption{(Color online) The mass-radius relation for neutron stars.}
\label{fig:6}
\end{figure}

In Fig.~\ref{fig:6} we present the mass-radius relation using the
EOS of the NL3 and TM1 models with both the weak and strong $YY$
interactions. It is shown that the results depend on the RMF
parameters used in the calculation. Since the pressure and density
inside neutron stars decrease from the center to the surface, the
most possible region where the deconfined quark phase can
exist is the center of the neutron star with maximum mass. We list in Table~%
\ref{tab:3} the properties of neutron stars with the maximum mass. It is found
that the central baryon density $n_{c}$ is between $n_{B}^{(1)}$ and $%
n_{B}^{(2)}$ for the NL3 model. This means that the neutron star can possess
a mixed phase core, but it is not dense enough to possess a pure quark core.
On the other hand, the values of $n_{c}$ in the TM1 model are smaller than
$n_{B}^{(1)}$, which means that the neutron star is only composed of hadronic
matter. As can be seen in Table~\ref{tab:3}, the neutron star properties
significantly depend on the RMF parameters used.


\section{Conclusions}

\label{sec:6}

We have studied the hadron-quark phase transition at high density,
which may occur in the core of massive neutron stars. In the present
work, we have adopted the RMF theory to describe the hadronic matter
phase, while the NJL model has been used for the quark matter phase.
With a definite EOS for the quark phase, we examine the influence of
the hadronic EOS on the deconfinement phase transition and neutron
star properties. In this paper, we employ two successful parameter
sets of the RMF model, NL3 and TM1, which have been widely used for
the description of nuclear matter and finite nuclei including unstable
nuclei. For each parameter set of the nucleonic sector,
we consider two cases of hyperon-hyperon interactions,
the weak and strong $YY$ interactions.
The hadron-quark phase transition can proceed through a mixed phase
of hadronic and quark matter over a finite range of pressures and
densities according to the Gibbs criteria for phase equilibrium. We
have found that the mixed phase starts at $n_{B}^{(1)}\simeq 0.50\
\mathrm{{fm}^{-3}}$ ($0.71\ \mathrm{{fm}^{-3}}$) and ends at
$n_{B}^{(2)}\simeq 0.75\ \mathrm{{fm}^{-3}}$ ($0.98\
\mathrm{{fm}^{-3}}$) for the NL3 parameter set with the weak
(strong) $YY$ interaction, while $n_{B}^{(1)}\simeq 1.31\
\mathrm{{fm}^{-3}}$ ($1.75\ \mathrm{{fm}^{-3}}$) and
$n_{B}^{(2)}\simeq 2.03\ \mathrm{{fm}^{-3}}$ ($2.49\
\mathrm{{fm}^{-3}}$) have been obtained for the TM1 cases. It is
shown that the hadron-quark phase transition occurs at lower
densities in the NL3 model than in the TM1 model, and in general the
weak $YY$ interaction leads to an earlier appearance of the mixed
phase. By comparing the results with different parametrizations in
the RMF model, we can see how sensitive the deconfinement phase
transition is to the hadronic EOS used in the calculation.

We have calculated the properties of neutron stars using the EOS over a wide
density range. The star properties such as their masses and radii are mainly
determined by the EOS at high density. We have found the maximum mass of neutron
stars falls in the range $1.62\sim 2.05\ M_{\odot }$ for the RMF parameters
used. With the NL3 model, the mixed phase can exist in the core of massive
neutron stars, but no pure quark phase can exist. For the TM1 model, the
neutron star is not dense enough to possess the mixed phase, and therefore
the hadron-quark phase transition could not occur inside neutron stars
in this approximation. We conclude that the maximum mass and
composition of neutron stars depend on the hadronic EOS adopted in the
calculation.

It is very interesting to compare our results with those previously published
in the literature~\cite{PRC99,NPA00,Menezes,Sharma}. In Ref.~\cite{PRC99},
the authors studied the possible hadron-quark phase transition utilizing the
same NJL model as used in the present work to describe the deconfined
quark phase, while the hadronic phase was described by several RMF models
(TM1, TM2, GL85, and GPS as listed in Table I of Ref.~\cite{PRC99}).
Comparing with our cases, different hyperon potentials were used to
constrain the hyperon couplings, but some of their values are not supported
by recent experimental observations. They found that the use of the GPS model
for the hadronic phase leads to the onset of the mixed phase at the energy
density $\epsilon \approx 7 \epsilon_0$ ($\epsilon_0=140\;\mathrm{MeV/fm^3}$
is the normal nuclear energy density), while the mixed phase does not appear
below $\epsilon \approx 10 \epsilon_0$ for the TM1, TM2, and GL85 models.
They concluded that within the model constructed in their calculation
the appearance of deconfined quark matter in the center of neutron stars
turns out to be very unlikely. This is consistent with our results although
the hyperon couplings are different between our study and theirs.
In Ref.~\cite{NPA00}, the authors employed an extended MIT bag model to describe
the quark phase and four RMF models (TM1, TM2, GL85, and GPS) to describe
the hadronic phase. They studied the influence of different hadronic EOS
and the influence of the model parameters of the quark phase on the properties
of the phase transition. Using the MIT bag model, they found the mixed phase
could appear below $\epsilon \approx 2 \epsilon_0$, which is much lower than
the threshold density obtained using the NJL model. This can be seen by
comparing Fig.~4 of Ref.~\cite{NPA00} with Fig.~9 of Ref.~\cite{PRC99}
and Fig.~3 of the present paper. In Ref.~\cite{Menezes}, the authors
investigated the deconfinement phase transition at zero and finite temperature
using both the MIT bag model and the NJL model for describing the quark phase
and a RMF model (GL model) for the hadronic phase. For the NJL model, however,
they adopted another parameter set which favors an earlier onset of the mixed
phase than the parameter set used in our calculation. For the hyperon couplings,
they used three choices and verified that for the NJL model the onset of
the mixed phase is very sensitive to the choice of the hyperon couplings.
Using the NJL model for the quark phase, they obtained two different behaviors.
One is the hyperons appear before the quarks and the mixed phase occurs at much
higher densities ($\sim 5 \rho_0$ with the normal nuclear matter density
$\rho_0=0.153\;\mathrm{{fm}^{-3}}$).
The other is the quarks appear at lower densities than the hyperons and
the mixed phase occurs at $\sim 2 \rho_0$. For almost all EOS used in their
study, the central energy density of the maximum-mass neutron star falls inside
the mixed phase, so the star can contain a core constituted by a mixed phase,
but it is not dense enough to possess a pure quark core.
In Ref.~\cite{Sharma}, the authors investigated the deconfinement phase
transition using an effective-field-theory-motivated RMF (E-RMF) model for the
hadronic phase and considering both unpaired quark matter (UQM) described by
the MIT bag model and paired quarks described by the color-flavor locked (CFL)
phase for the quark phase. They mentioned that they could not get any
mixed phase with the original G2 parameter set in the E-RMF model, so they
changed incompressibility $K$ from $215$ to $300$ MeV and the effective mass
$m_N^*/m_N$ from 0.664 to 0.7 to determine a modified parameter set G2$^*$
as given in Table I of Ref.~\cite{Sharma}. For the hyperon couplings,
they assumed that all the hyperons in the octet have the same couplings.
Using the G2$^*$ parameter set in the E-RMF model for the hadronic phase
and $B^{1/4}=170\;\mathrm{MeV}$ in the UQM model for the
quark phase, they obtained the onset of the mixed phase at $\sim 1.3 \rho_0$,
while the appearance of the mixed phase starts at $\sim 2.4 \rho_0$
using the CFL model with $B^{1/4}=188\;\mathrm{MeV}$ for the quark phase.
In our cases, the mixed phase occurs at $\sim 3.3 \rho_0 $ ($8.6 \rho_0$)
using the NL3 (TM1) model with the weak $YY$ interaction for the hadronic phase
and the NJL model for the quark phase. For simplicity, we have not considered
the color-flavor locked phase in the present work.
By comparing all these results, we conclude that the deconfinement phase
transition is very sensitive to both the hadronic EOS and the quark EOS,
whether or not the deconfined quark matter appears in the center
of neutron stars depends on the models adopted in the calculation.


\section*{Acknowledgments}

This work was supported in part by the National Natural Science Foundation
of China (No. 10675064) and the Specialized Research Fund for the Doctoral
Program of Higher Education (No. 20040055010).


\newpage
\begin{table}[tbp]
\caption{The parameter sets NL3~\cite{NL3} and TM1~\cite{TM1} used
in the calculation. The masses are given in MeV.}
\label{tab:1}
\begin{center}
\begin{tabular}{lccccccccccc}
\hline\hline  & $m_N$ & $m_\sigma$ & $m_\omega$ & $m_\rho$ &
$g_{\sigma N}$& $g_{\omega N}$ & $g_{\rho N}$ & $g_2$ ($\mathrm{{fm}^{-1}}$) & $g_3$ & $c_3$  \\
\hline
NL3 &  939.0 & 508.194 & 782.501 & 763.0 & 10.217 & 12.868 & 4.474 & -10.431 & -28.885 & -       \\
TM1 &  938.0 & 511.198 & 783.0   & 770.0 & 10.029 & 12.614 & 4.632 & -7.233  & 0.618   & 71.308  \\
\hline\hline
\end{tabular}
\end{center}
\end{table}

\begin{table}[tbp]
\caption{The scalar coupling constants for the hyperons.}
\label{tab:2}
\begin{center}
\begin{tabular}{lccccc}
\hline\hline    & $g_{\sigma\Lambda}$ & $g_{\sigma\Sigma}$ &
$g_{\sigma\Xi}$ & $g_{\sigma^{\ast}\Lambda(\Sigma)}$ &
$g_{\sigma^{\ast}\Xi}$ \\ \hline NL3 (weak $YY$) & 6.269 & 4.709 & 3.242 & 5.595 & 11.765 \\
NL3 (strong $YY$) & 6.269 & 4.709 & 3.242 & 7.138 & 12.809 \\
TM1 (weak $YY$) & 6.170 & 4.472 & 3.202 & 5.412 & 11.516  \\
TM1 (strong $YY$) & 6.170 & 4.472 & 3.202 & 7.018 & 12.600 \\
\hline\hline
\end{tabular}
\end{center}
\end{table}

\begin{table}[tbp]
\caption{The properties of neutron stars with the maximum mass $M_{\mathrm{max}}$.
The central energy density, pressure, and baryon number density are denoted
by $\varepsilon_c$, $P_c$, and $n_{c}$, respectively. $R$ and
$R_{\mathrm{MP}}$ denote the radii of the star and its mixed phase
core.}
\label{tab:3}
\begin{center}
\begin{tabular}{lcccccc}
\hline\hline & $M_\mathrm{max}$ & $\varepsilon_c$ & $ P_c$ &
$n_c$ & $R$ & $R_\mathrm{MP}$   \\
&$(M_\odot)$ & $(10^{15}\, \mathrm{{g}/{cm}^3) }$ & $(10^{35}\,
\mathrm{{dyn}/{cm} ^2) }$ & $(\mathrm{{fm}^{-3})}$ & (km) &
(km)  \\
\hline
NL3 (weak $YY$)   & 2.02 & 1.42 & 2.08 & 0.69 & 13.80 & 4.21 \\
NL3 (strong $YY$) & 2.05 & 1.46 & 2.62 & 0.71 & 13.48 & 0.31 \\
TM1 (weak $YY$)   & 1.71 & 1.50 & 2.08 & 0.75 & 13.07 & -\\
TM1 (strong $YY$) & 1.62 & 1.23 & 1.32 & 0.63 & 13.62 & -\\
\hline\hline
\end{tabular}
\end{center}
\end{table}

\newpage

\end{document}